\documentstyle[prl,aps,psfig,floats,twocolumn]{revtex} 
\newcommand{\avg}[1]{\langle{#1}\rangle}
\newcommand{\req}[1]{(\ref{#1})}
\newcommand{\beq}{\begin{equation}}
\newcommand{\beqar}{\begin{eqnarray}}
\newcommand{\eeqar}{\end{eqnarray}}
\newcommand{\beqars}{\begin{eqnarray*}}
\newcommand{\eeqars}{\end{eqnarray*}}
\newcommand{\eeq}{\end{equation}}
\begin{document}
\draft
\twocolumn[\hsize\textwidth\columnwidth\hsize\csname 
@twocolumnfalse\endcsname
\title{A Prototype Model of Stock Exchange}

\author{G. Caldarelli}
\address{Istituto Nazionale di Fisica
della Materia (INFM), Sezione di Trieste and SISSA Trieste,}
\address{Department of Theoretical Physics, The University,
M13 9PL Manchester, U.K.}
\author{M. Marsili, Y.-C. Zhang}
\address{Institut de Physique Th\'eorique, Universit\'e de Fribourg
P\'erolles, Fribourg CH-1700}

\date{14-02-1997}
\maketitle
\begin{abstract}
A prototype model of stock market is 
introduced and studied numerically. In this self-organized
system, we consider only the interaction among traders without 
external influences. Agents trade according to 
their own strategy, to accumulate his assets by speculating on the 
price's fluctuations which are produced by themselves.
The model reproduced rather realistic price histories 
whose statistical properties are also similar to those 
observed in real markets.
\end{abstract}

\pacs{SISSA Ref 22/97/CM}
]
\narrowtext

In the modern market of stocks, currencies, and commodities, trading 
patterns 
are becoming more and more global.
Market-moving information is being transmitted quickly to all 
the participants (at least in principle). However, not all 
the participants interpret the information the same way and 
react at the same time delay. In fact, every participant has 
a certain fixed framework facing external events. It is well 
known that the global market is far from being at 
equilibrium\cite{equil}, the collective behavior of the market can 
occasionally have violent bursts (rallies or crushes) and 
these violent events follow some empirically well established scaling 
laws. 
These are currently the subject of intensive studies 
\cite{man,ms,ab,mf,gcmz}. It is not settled yet
whether these fluctuations are due to external factors or
to the inherent interaction among market's players.

From a physicist's point of view, the market is an excellent 
example of self-organized systems: each agent decides 
according to his own perception of the events. 
In the simplest framework, these events consist in the price 
fluctuations, the only available information. 
Each participant's action will in turn influence the price. 
In a true economy there are external driving factors, such 
as politics, natural disasters, human psychology etc. 
Another systematic effect is due to the periodicity of 
human life (days, weeks, months and years) which also 
influences the dynamics of prices. From the theoretical side, it
is interesting to understand whether the statistical properties
of prices depend directly on the external driving factors
or whether they are self generated by the system itself.

In the present work, sticking to a physicist's point of view, 
we shall address this issue by investigating this system in the 
absence of external factors. Thus in our market all
the participants are speculators: they trade with the sole
aim to increase their capital. We shall see that a very rich
and complex statistics of price fluctuations emerges from 
such a closed
system of traders which speculate on the price fluctuations
they produce themselves. In spite of the simplicity of our
model and of the strategies of the single participants, and
the outright exclusion of economic external factors, we shall
find a market which behaves surprisingly realistically.
These results suggest that  a stock market can be 
considered as a self-organized critical system: The system 
reaches dynamically an equilibrium state characterized by 
fluctuations of any size, without the need of any parameter 
fine tuning or external driving.

Let us define our model more precisely. Each player is 
initially given the same amount of capital in two forms: 
cash $M_{i,t=0}$ and stock $S_{i,t=0}$. At any time $t$ the capital 
of player $i$ is
given by $C_{i,t}=M_{i,t}+p_tS_{i,t}$, where $p_t$ is the current 
price of the stock. There is only one stock in this model, 
e.g. a foreign currency.
All trading consists of switching back and forth between 
cash and this stock. Each player has a strategy that makes 
recommendation for buying or selling a certain amount of stock for the 
next time step. This depends solely on the information available,
i.e. the past price history. All the 
players have equal access to the price history. 
The actions taken by each player is bounded by
his belongings. Player $i$ can invest only a fraction
of his stocks which, at any time $t$, is given by 
his strategy:
At time $t$, 
the general form of the strategy of player $i$ is
\[{X}_{i,t}=F_i[p_t,p_{t-1},\ldots]\]
where $X_{i,t}S_{i,t}$ is the amount of stock player
$i$ decides to buy ($X_{i,t}>0$) or to sell ($X_{i,t}<0$). 
Our model draws inspiration 
from Brian Arthur's model of Bar Attendance \cite{bar}, 
where each bar hopper can formulate his own prediction, 
based on the past observation. This shows that, in
a game of interacting strategies, the measure of 
a strategy's efficacy can be given only {\em a posteriori}.
Any strategy is, {\em a priori}, as good as any other.
Therefore, in our case, initially the strategies are 
randomly chosen. Then, at each time step, the agent with 
the smallest capital is eliminated and replaced by one 
with a new (random) strategy.
This refreshing rule keeps the population of
the traders dynamic and it is a simple application of
Darwinism to economy. 

Since, the ``space of strategies'' is enormous,
finding some ``local maximum'' of the fitness is nearly impossible.
Moreover it is unrealistic to assume that the action
of player $i$ is independent of his belongings
$M_{i,t}$ and $S_{i,t}$. 

For these two reasons we {\em i)} parametrized the
functions $F_i$ in terms of indicators $I_k\{p\}$
and {\em ii)} we introduced a restoring effect
which tends to balance the ratio $S_{i,t}/M_{i,t}$ to the 
current price $p_t$. 
For the indicators we choose moving time averages 
of combinations of time derivatives of $\log p_t$ 
(e.g. $I_1=\avg{\partial_t \log p_t}$, 
$I_2=\avg{\partial_t^2 \log p_t}$,
$I_3=\avg{[\partial_t \log p_t]^2}$, etc). The time
averages were done over a time period of typically
$10\sim 100$ time steps\cite{nota}. 
Note that considering time differences
of $\log p$ and not of $p_t$, makes the indicators,
and hence the strategies, depend on relative
fluctuations of $p_t$ and not on his absolute value.
The strategies were then parametrized by $\ell$ numbers
$\eta_{i,k}$:
\[{X}_{i,t}=f\left(\sum_{k=1}^\ell \eta_{i,k} 
I_k\{p\}\right)\]
where $f(x)$ is a non-linear function. The nonlinearity of $f(x)$ is
introduced to mimic, to some extent, the behavior of agents
in a realistic market. Agents indeed sometimes play ``contrarian''
strategies, i.e. strategies which do not follow the trend.
In a rally run, an agent may take advantage of short selling 
in anticipation of a reverse or crush. 
Furthermore, $f(x)$ must be less than $1$, since it 
represents the fraction of one agent's stock moved in a time 
step. Note also that large arguments of $x$ occur for example in 
the presence of wild fluctuations of $p_t$. The behavior
of traders becomes in these cases very cautious, which
implies $f(x)\ll 1$ for $x$ large. 
To imitate these behaviour we took $f(x)=x/[1+(x/2)^4]$. 

The amount of stock $\Delta S_{i,t}$ agent $i$
decides to sell ($\Delta S_{i,t}<0$) or to buy ($\Delta S_{i,t}>0$)
is given by
\beq
\Delta S_{i,t}={X}_{i,t}S_{i,t}+\frac{\gamma_i 
M_{i,t}-p_tS_{i,t}}{2\tau_i}.
\label{ds}
\eeq
Here the first term is pure speculation, whereas the
second introduces a dependence on $M_{i,t}$ and $S_{i,t}$
in the action of player $i$. 
In order to motivate this term, consider the event where,
for some reason the price remains constant $p_t=p_0$ for a
long period. 
A realistic behavior of the agents in such a situation 
would be that to ri-equilibrate their portfolio 
at their chosen level $S_{i,t}/M_{i,t}=p_0/\gamma_i$. 
These operations 
at constant price do not change the capitals $C_{i,t}$.
On the contrary, having his assets equilibrated
at the actual price, a player is in the most favorable situation 
to face possible future price fluctuations of either 
sign. The second term in eq. \req{ds} reproduces this behavior. 
Indeed if the price is constant, all indicators $I_k$ 
vanish and so do all speculation terms $X_{i,t}$. 
In the absence of the first term, the second equilibrates 
the ratio $S_{i,t}/M_{i,t}$ to a value $p/\gamma_i$ within a 
time of order $\tau_i$.

In summary the strategy of each player is 
parameterized by $\ell+2$ numbers: $\eta_{i,1},\ldots,
\eta_{i,\ell},\gamma_i$ and $\tau_i$.
Once the price $p_t$ is fixed, it is communicated to 
the players who can decide their actions $\Delta S_{i,t}$. 
The transactions then take place at this price $p_t$.
Let us define the total demand and offer of stocks at time $t$ 
\[D_t=\sum_{i: \Delta S_{i,t}>0} \Delta S_{i,t}\qquad
O_t=-\sum_{i: \Delta S_{i,t}<0} \Delta S_{i,t}.\]
When the demand is larger than the offer, the
players willing to buy $\Delta S_{i,t}>0$ stocks
will in fact be able to buy only the available
amount $\bar\Delta S_{i,t}=\Delta S_{i,t} O_t/D_t$, whereas
players who sell will sell all their $-\Delta S_{i,t}$ 
stocks ($\bar\Delta S_{i,t}=\Delta S_{i,t}$).
The reverse situation will clearly apply if $D_t<O_t$.
Our model also includes a finite cost $\pi
\Delta S_{i_t}$ on the buyers ($\pi\sim 10^{-3}$) and 
random fluctuations $\epsilon_{i,t} M_{i,t}$ in the 
cash of each agent ($\epsilon_{i,t}$ is a random 
variable uniformly distributed in $[-\epsilon,\epsilon]$)
which, loosely speaking, represents the ``heat bath''
fluctuations due to all the other actions of player $i$.
These rules are summarized in the following equation:
\beqars
\left.
\begin{array}{l}
  S_{i,t+1}=S_{i,t}+\bar\Delta S_{i,t}\\
  M_{i,t+1}=[1+\epsilon_{i,t}][M_{i,t}-p_t(1+\pi)\bar\Delta S_{i,t}]
\end{array}
\right\}&~&\hbox{if $\bar\Delta S_{i,t}>0$},\\
\left.
\begin{array}{l}
  S_{i,t+1}=S_{i,t}+\bar\Delta S_{i,t}\\
  M_{i,t+1}=[1+\epsilon_{i,t}][M_{i,t}-p_t\bar\Delta S_{i,t}] 
\end{array}
\right\}&~&\hbox{if $\bar\Delta S_{i,t}<0$}.
\eeqars

\begin{figure}
\centerline{\psfig{file=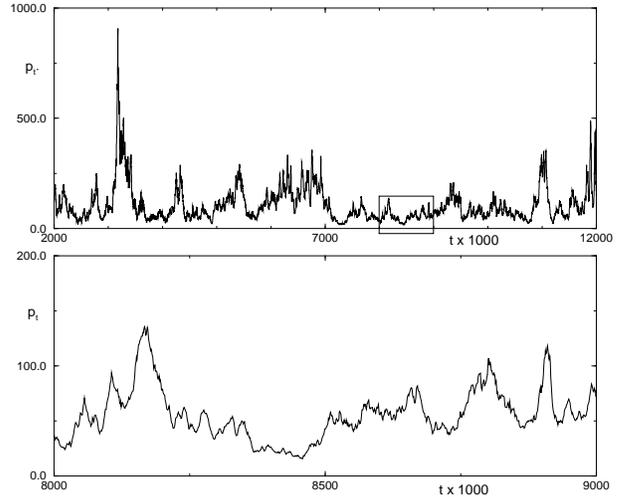,width=8.5cm,angle=270}}
\caption{Price history for a system of $1000$ agents.
The parameters are $\epsilon=0.01$ and $\pi =10^{-3}$.
In the lower part the zoom of the area in the upper rectangle.}
\label{p_t}
\end{figure}

Next a new price $p_{t+1}$ is determined. 
This is done implementing the law
of demand and offer in the form
\beq
p_{t+1}=p_t\frac{\avg{D_t}}{\avg{O_t}}
\label{domoff}
\eeq
where the averages are, as before time averages. Note that
the price raises when there is a large demand and falls
when the offer is large. Also note that this form of 
the law of demand and offer is dimensionally correct.

Our numerical simulations are quite encouraging. Despite the 
simplicity and the arbitrariness of the strategies, 
an extremely rich price history is created. 
A sample of $p_t$ is shown in fig. \ref{p_t}. 
This shows fluctuations of all sizes. 
Depending on the parameters, on long 
runs, also some crushes occasionally occur, with almost no 
sign of its coming. These crushes arise only as a result of 
collective trading activity. Apart from a simple ``smoothing''
\cite{nota} which becomes more efficient in the presence
of wild fluctuations, our model does not implement
the many corrections which are taken in similar cases by 
central authorities in a real market.

The signal is similar to that of stock prices or foreign 
currencies. This impression is confirmed analyzing 
the temporal signal. Recently quite a few empirical 
studies have been carried out for the statistics of economics 
time series, notably the Standard \& Poor's 500 index\cite{ms}, 
and high frequency foreign exchange data\cite{mf,gcmz}. 

When comparing our statistical data, the definition of time 
becomes a matter of concern. In our model the time flows 
uniformly and the number of agents are fixed. In a real economy 
there are periods of inactivity, when the market closes, and 
the number of active agents varies with time. This leads to
systematic periodic variations in the signal $p_t$ so that
its fluctuations $x=p_{t+\tau}-p_t$ can no
longer be considered as a stationary variable. This issue 
was discussed at length in ref. \cite{gcmz} where a time 
transformation which eliminates these systematic effects was 
introduced. 
Our model clearly does not contain these 
systematic effects and therefore better compares with real 
economic data in the transformed time\cite{gcmz}.

\begin{figure}
\centerline{\psfig{file=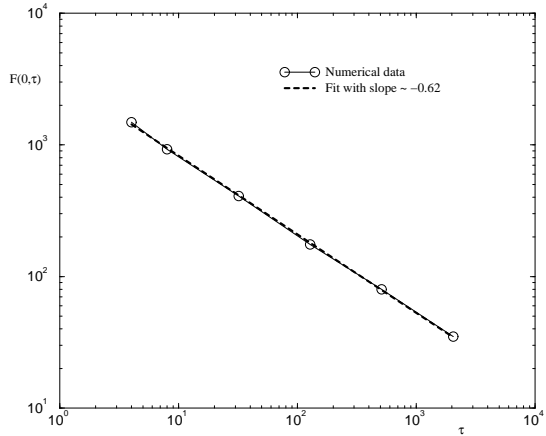,width=8.5cm,angle=270}}
\caption{Probability $F(x=0,\tau)$ of price returns 
(i.e. $x=0 \Rightarrow p_{t+\tau}=p_t$) vs $\tau$.
The parameters are the same as in fig. 1.}
\label{dp2}
\end{figure}

Following refs. \cite{ms,mf,gcmz}, let us define the 
histogram $F(x,\tau)$ of price variations $x=p_{t+\tau}-p_t$.
Figure \ref{dp2} shows that the scaling behavior of the
the price ``returns'' $F(0,\tau)$ is very similar to that observed
in a real economy, it behaves like $\tau ^{-H}$ with an exponent
$H \simeq 0.62$.

\begin{figure}
\centerline{\psfig{file=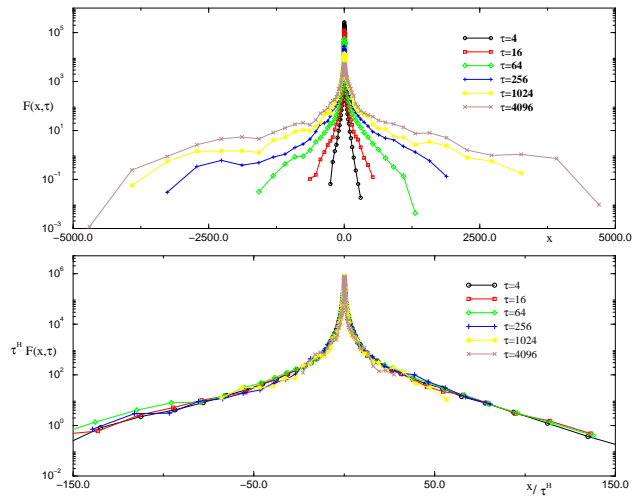,width=8.5cm,angle=270}}
\caption{Upper part: histogram of price variations for several
values of $\tau$. Lower part: collapse plot 
of the previous data with $H=0.62$.
The parameters are the same as in fig. 1.}
\label{dist}
\end{figure}

The full distribution $F(x,\tau)$ is shown in figure \ref{dist}
for several values of $\tau$. These distributions satisfy the
scaling hypothesis
\begin{equation}
F(x,\tau)= \tau^{-H}F\left(x\tau^{-H},1\right)=
\tau^Hg\left(x\tau^{-H}\right),
\label{scaling}
\end{equation}
with an exponent $H = 0.62$,
as shown in the lower part of fig. \ref{dist}.
This is very similar to the behavior of prices in
a real economy. The tails of the distributions 
have a power law character $F(x,\tau)\sim x^{-\alpha}$ 
with exponent $\alpha$ close to $2$. 
This differs from what is seen in real
economic data where either the exponent is larger ($\approx 
4.5$ \cite{gcmz}) or the decay is more rapid \cite{ms}.
We believe that this is due to the fact that the tail of the 
distribution describes extreme events. Under extreme events, 
in a real economy, the rules of the game change drastically. 
On the contrary in our model the rules are always the same 
no matter what fluctuations $p_t$ may suffer.

\begin{figure}
\centerline{\psfig{file=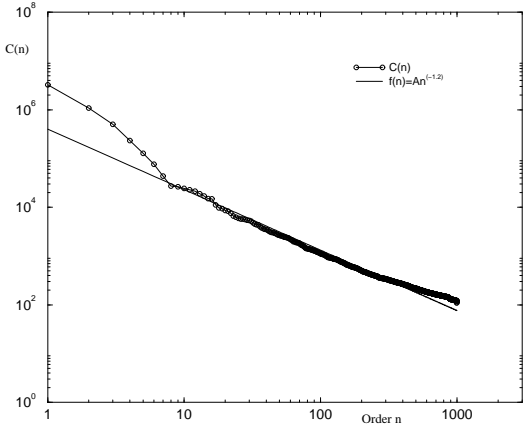,width=8.5cm,angle=270}}
\caption{The capitals $C(n)$ owned by each player in order of richness.
The parameters are the same as in fig. 1.}
\label{zipf_m}
\end{figure}

In the stationary state, one can classify the traders 
according to their wealth. Zipf \cite{zipf}
has observed that the distribution of economic power among
individuals of a society follow a well defined power law, 
hence the name Zipf's law. In our artificial 
society we plot the assets distribution in the Zipf's fashion
where the capital $C(n)$ of the $n^{\rm th}$ richer person is 
plotted against the order $n$.
Figure \ref{zipf_m} shows this law (note that 
the plot refers to one single snapshot of a system of 
$1000$ agents). Our exponent $\approx -1.2$ is not too far from 
Zipf's value for the distribution of cooperation assets\cite{zipf}.

With respect to the robustness of the results, 
we found that particular care has to be paid in order to
avoid crushes and singularities in the market. 
For all choices of parameters yielding a stable
behavior, we found similar results. 
But, for example, including ``copying'',i.e. the possibility for 
a poor to copy the strategy of a richer, changes the exponent 
$H$ from $0.62$ to $0.5$. Interestingly, in the model 
with copying, we also detected a multiscaling behavior of 
the same form of the one discussed in \cite{mf}. A weak
multiscaling, or no multiscaling at all, was instead found
in the dynamics without copying. 

In conclusion we find realistic behavior in a simple model 
for a stock market without external driving.
The players trade in an non-ending fight 
against each other to survive. Such a dynamic
system produce a non trivial time series behind, which records all the 
infighting and ingenuity of the players trying to out-guess others.
Simple signal forms are expected to be excluded
since they are too easy to anticipate. 
The result is a signal with all the
surprises of all sizes (for the traders as well as for us!).
We have purposely excluded any real information input, our results
nevertheless show close resemblance to real markets. This suggests that
the statistics we observe in real markets is 
mainly due to the interaction among ``speculators" trading on
technical grounds, regardless of economic fundamentals.
Indeed it is known that in Foreign exchange markets the trades by 
speculators far outnumbers the trades due to real commercial needs.

It is clear that by no means one can conclude that our 
model captures all the relevant aspects of a real market. 
As already said it misses the effects of external drive. More
importantly, it does not contain adaptive dynamics of the 
player's strategies. 
Further studies should answer the question: what are the essential 
elements in a model that will reproduce realistic results? 
In this perspective our model can be considered as a first 
step in this challenging direction.

GC acknowledges the University of Fribourg for the kind 
hospitality, Anna B. and Domenico C. for useful hints. 
We are also grateful to S. Galluccio for helpful discussions.


\begin{thebibliography}{10}
\bibitem{nota} Time averages were computed run time using an 
exponential kernel: $\avg{A}_{t}=(A_t+\lambda\avg{A}_{t-1})/
(1-\lambda)$. The value of $\lambda$ was also determined run 
time according to the volatility: During periods with large
fluctuations of $p_t$ the time average was extended (i.e. 
$\lambda$ was increases towards $1$) to a longer period.
\end{thebibliography}

\end{document}